\documentclass[twocolumn,showpacs,showkeys,superscriptaddress]{revtex4}

\usepackage{bm}
\usepackage{color}
\usepackage{amsmath}
\usepackage{natbib}

\begin{document}

\title{Diamagnetic  field states in cosmological plasmas}

\author{Felipe A. Asenjo}
\email{felipe.asenjo@uai.cl}
\affiliation{Facultad de Ingenier\'{\i}a y Ciencias, Universidad Adolfo Ib\'a\~nez, Santiago 7941169, Chile.}
\author{Swadesh M. Mahajan}
\email{mahajan@mail.utexas.edu}
\affiliation{Institute for Fusion Studies, The University of Texas at Austin, Texas 78712, USA.}

\date{\today}

\begin{abstract}
Using a generally covariant Electro-Vortic (magnetofluid) formalism for relativistic plasmas, the dynamical evolution of a generalized vorticity (a combination of the magnetic and kinematic parts) is studied in a cosmological context. 
We derive  macroscopic vorticity and magnetic field structures that can emerge in spatial equilibrium configurations of the relativistic plasma. These fields, however, evolve in time.
These  magnetic and velocity fields fields are self-consistently sustained in a diamagnetic state in the expanding Universe,
and do not require an external  seed for their existence. In particular, we explore a special class of magnetic/velocity field structures supported by a plasma in which the generalized vorticity vanishes. We derive a highly interesting characteristic of such  ``superconductor--like" fields  in a cosmological plasmas in the radiation--era in early Universe. In that case,  the fields grow proportional to the scale factor, establishing a deep connection between the expanding universe and  the primordial magnetic fields.
\end{abstract}

\pacs{04.70.Bw, 52.27.Ny, 52.35.We, 95.30.Qd}


\maketitle


\section{Introduction}

Exploring the interaction of gravitational fields and inhomogeneous plasma thermodynamics as a possible source of primordial magnetic fields has, recently, received considerable attention \cite{asenjo1,asenjo2,chinmoy1,chinmoy2,chinmoy3,massimo1,massimo2,khanna}. Much of this work has been carried out within the framework of what has been called a unified magneto--fluid, recently generalized to, Electro-Vortic (EV) formalism \cite{mah 03, mahajanElectroVortical}. The primary new construct of this formalism is the EV tensor $M^{\mu\nu} = F^{\mu\nu}+(m/q) S^{\mu\nu}$, a weighted sum of the Electromagnetic $F^{\mu\nu}$ and the Vortical $S^{\mu\nu}$ field tensors; the latter representing both the kinematic and thermal content of the relativistic hot fluid. In the EV formalism, the fluid dynamics  reduces to a simple Helmholtz vortical form in terms of new composite variables, the most familiar being the so called generalized vorticity $\bm\Omega$ (such that $\Omega^{i}=\epsilon^{ijk}{M}_{jk}$) that has  both magnetic and thermal-kinetic parts. The most important message is that in the considerably complicated dynamics of a hot relativistic fluid, $\bm\Omega$ plays the same role as the magnetic field $\bm B$ does in the much simpler magnetohydrodynamics (MHD). Since many of our familiar concepts emerged from an MHD description, it is important  to note the connections between EV dynamics and MHD. 

The earlier special relativistic theory  \cite{mah 03} was extended  to explore general relativistic effects in Refs.~\cite{asenjo1,asenjo2,chinmoy1,chinmoy2,chinmoy3}. It was shown that a combination of gravity modified Lorentz factor of the fluid element and the spatial variation of plasma thermodynamics, leads to an additional (to the special relativistic mechanism \cite{mah1,mah12}) source that  creates a vorticity seed out of a state with no initial vorticity. This can be viewed as a grand generalization of the MHD or Biermann battery where a magnetic field seed is provided through baroclinic thermodynamics.
The seed vorticity (and therefore a magnetic field), then, can be amplified, for example,  by a dynamo mechanism. These calculations have shown that that a plasma around a Kerr black hole can produce a larger magnetic field seed  than a corresponding Schwarzschild system \cite{chinmoy2}.
In a different approach, based on  the gradient expansion formalism for a cosmological spacetime geometry, the kinematic vorticity and magnetic fields in an electron-ion-photon pre-decoupled plasma have been studied \cite{massimo2}. The two formalisms are different in that the former explores a unified formulation in which it is possible to construct a conserved (in the absence of the drive) generalized vorticity (combining  the magnetic and the kinematic); they also  treat plasma thermodynamics differently. The latter does not consider  relativistic translations  of the classical  Biermann battery.

It is well known that a relativistic plasma can also sustain self-consistent equilibrium configurations, states that represent self-consistent macroscopic structured magnetic and flow fields without any externally generated seeds (see Refs.~\cite{mahajanclassicaldiag,mahajanElectroVortical,MahajanYoshi} and references therein). In particular, some of these equilibrium states resemble the features of a superconductor as the plasma displays perfect diamagnetism.

In all these relativistic investigations, the gravitational field was specified as a static background (the space time metric is independent of time). Such calculations do capture essential features, for instance, of what may be happening in plasmas in the accretion disks around compact objects. However, a purely static approach is not enough when applied to the universe as a whole because the expansion of the universe (a time dependent metric) is a fundamental  element of cosmology that must be contended with. In this work, we reinvestigate the primordial magnetic field problem in an expanding universe described by  the Friedmann-Robertson-Walker (FRW) metric.

Our aim is to seek self-consistent magnetic-velocity field solutions (in the spirit of  \cite {mahajanElectroVortical}) that evolve with the universe through an explicit dependence on the scale factor. Interestingly enough we find that the magnetic field strength increases with the scale factor, at least partially compensating the automatic dilution caused by the expanding universe.  
      
The paper is organized as follows. In Sec. II, we  review some of the essentials of the the ElectroVortic formalism. In Sec. III, we develop the 3+1 formalism used to describe the  cosmological fluid, and  derive the  dynamical equation for the generalized vorticity.   In Sec. IV and V, we study time varying  spatial ``equilibrium" solutions with their spatial structure resembling that of a perfect diamagnetic plasma state. The results are discussed  in Sec. VI.

\section{Unified plasma dynamics in curved spacetime}
\label{dynamics}

The dynamics of an ideal plasma (a charged fluid), immersed in an electromagnetic field $F_{\mu\nu}$, is contained in the conservation law 
\begin{equation}
\nabla_\nu{T^{\mu\nu}}=q n F^{\mu\nu}U_\nu\, ,
\label{GRenmomen}
\end{equation}
where $U^{\mu}$ (such that $U^\mu U_\mu=-1$) is the four velocity, $\nabla_\nu$ is the covariant derivative for the metric $g_{\mu\nu}$ describing a curved spacetimes ($c=1$), and
\begin{equation}
 T^{\mu\nu}=m n f U^\mu U^\nu+p g^{\mu\nu}\, , 
\label{GRenmomentensor}
\end{equation}
is the energy-momentum tensor for an ideal plasma \cite{mah 03,Bek}; the charge $q$ and the mass $m$ of the fluid element are scalar invariants. The definition of the energy-momentum \eqref{GRenmomentensor}
involves three thermodynamic scalars: the scalar number density $n$ ( the rest frame density),  the scalar pressure $p$,  and enthalpy density $h=m n f$, where $f $ is a function of temperature $T$ that in the special case of a relativistic
Maxwell distribution becomes $f= K_3(m /k_B T)/ K_2(m /k_B T)$ where $K_j$ is the modified Bessel functions of order $j$, and $k_B$ is the Boltzmann constant.
The system is completed with the continuity  
 \begin{equation}\label{conTnU}
\nabla_\mu\left(nU^\mu\right)=0\, ,
\end{equation}
and Maxwell equations
\begin{equation}
\nabla_\nu {F^{\mu\nu}}=4\pi q n U^\mu\, .
 \label{Maxwellcurved}
\end{equation}

The  global dynamics given by Eqs.~\eqref{GRenmomen}-\eqref{Maxwellcurved} is more conveniently studied in terms of a unified  field tensor  \cite{mah 03, asenjo1,Bek}, 
\begin{equation}\label{Mu}
M^{\mu\nu}=F^{\mu\nu}+\frac{m}{q}S^{\mu\nu}\ ,
\end{equation}
in which all kinematic and thermal (through $f$) aspects of the fluid are now represented by  the antisymmetric tensor \cite{mah 03, asenjo1,Bek}
\begin{equation} 
S^{\mu\nu}=\nabla^\mu\left(fU^{\nu}\right)-\nabla^\nu\left(fU^{\mu}\right)\, .
\end{equation}
The resulting equation of motion takes the form ($\partial^\mu$ is the four  derivative)
\begin{equation}
q\ U_{\nu} M^{\mu\nu}=-T\partial^\mu\sigma\, .
\label{eqmo}
\end{equation}
where $\sigma$ is the scalar entropy density of the fluid, and it is related to pressure through
\begin{equation}\label{entropyGrad}
\partial^\mu \sigma=\frac{mn \partial^\mu f-\partial^\mu p}{nT}\, ,
\end{equation}
We can see that due to the antisymmetry of $M_{\mu\nu}$, the fluid description presented here is always  isentropic 
\begin{equation}\label{consentrioydensi}
U_{\mu}\partial^\mu\sigma=0\, .
\end{equation}

This general formulation of plasma dynamics in curved spacetime allows even a more compact expression for a special class of velocity fields. If the four velocity were derivable from a Clebsch potetial Q, \cite{mahajanElectroVortical},  
\begin{equation}\label{ClebschU}
U^\mu=\frac{1}{T}\partial^\mu Q\, ,
\end{equation}
then \eqref{eqmo} reduces to
\begin{equation}
U_{\nu} {\cal M}^{\mu\nu}=0\, , 
\label{eqmoUM}
\end{equation}
where the new field tensor 
\begin{eqnarray}
{\cal M}^{\mu\nu}&=&{M}^{\mu\nu}-\frac{1}{q}\partial^\mu\left(\sigma\partial^\nu Q\right)+\frac{1}{q}\partial^\nu\left(\sigma\partial^\mu Q\right)\nonumber\\
&=&{M}^{\mu\nu}-\partial^\mu\left(\frac{\sigma T}{q} U^\nu \right)+\partial^\nu\left(\frac{\sigma T}{q}U^\mu\right)\, .
\end{eqnarray}
represents complete Electro-Vortic unification of the ideal relativistic dynamics. 

\section{Explicit Plasma Dynamics in cosmology}
\label{dynamicsCosmo}

For an explicit formulation of plasma dynamics in the the cosmological background, we introduce the  FRW 
metric. Since the focus of this calculation is to figure out how the plasma dynamics is affected by  the expansion of the Universe, we will restrict  ourselves  to a spatially flat universe; the corresponding FRW metric will be  \cite{ryden,misner}
\begin{equation}
ds^2=-dt^2+a^2\gamma_{ij}dx^idx^j\, ,~~~(i,j=1,2,3)
\label{eqcm}
\end{equation}
$a=a(t)$ is the time-dependent scale factor of the Universe, and $\gamma_{ij}=(1,1,1)$ is the 3-metric of the spacelike hypersurfaces of the flat spacetime. 

Following the procedure described in Refs.~\cite{asenjo1,asenjo2,chinmoy1,chinmoy2,thorne}, we perform a $3+1$ decomposition of the covariant fluid plasma equations ( Eq.~\eqref{eqmo} with the FRW metric) in order to put them in a more intuitive vectorial form.
 We define a normalized timelike vector field $n^\mu$, obeying $n^\mu n_\mu=-1$ and $n^\mu \gamma_{\mu\nu}=0$, such as $n_\mu=(1,0,0,0)$ and $n^\mu=(-1,0,0,0)$. The  projection into time-like and space-like hypersurfaces, then,  is readily achieved by contracting every tensor with $n^\mu$  and  $a^2\gamma_{\mu\nu}$.

We may write the four-velocity $U^\mu=(\Gamma, \Gamma v^i)$ 
\begin{equation}
U_\mu=-\Gamma n_\mu+a^2\Gamma \gamma_{\mu\nu} v^\nu\, ,
\label{velocityU3+1}
\end{equation}
where $v^i=dx^i/dt$ corresponds to the spatial $i$-component of the fluid velocity $\bm{v}$.
It follows that $n_\mu U^\mu=\Gamma$, where the Lorentz factor is given by
\begin{equation}
\Gamma=\left(1-a^2{v}^2\right)^{-1/2}\, ,
\label{gammacurved}
\end{equation}
where ${v}^2=\gamma_{ij}v^i v^j={\bm v}\cdot{\bm v}$. Note that the scale factor modifies $\Gamma$ from its special relativistic value. Using the 3+1 decomposition \eqref{velocityU3+1} of the FRW metric, the continuity equation \eqref{conTnU} becomes
\begin{equation}
\frac{1}{a^3}\frac{\partial}{\partial t}\left(a^3 n \Gamma\right)+\nabla\cdot \left(n\Gamma {\bm v}\right)=0\, ,
\end{equation}
where $\nabla$ is the flat spatial gradient operator.

Now, neglecting the plasma
back-reaction on spacetime, one may write down the decomposition of the field equations. The $3+1$ decomposed \cite{pla1,pla3,pla4,pla5,pla2,pla6,pla7,pla8,tajima,tarkenton,acht} electric and magnetic fields are obtained as
\begin{equation}
 E^\mu=n_\nu F^{\nu\mu}\, ,\qquad B^\mu=\frac{1}{2}n_\rho \epsilon^{\rho\mu\sigma\tau}F_{\sigma\tau}\, ,
\label{EBdecomposed}\end{equation}
where $\epsilon^{\alpha\beta\gamma\delta}$ is the totally antisymmetric tensor. Notice that the electric $E^\mu$ and the magnetic $B^\mu$ fields are spacelike tensors ($n_\mu E^\mu=0$ and $n_\mu B^\mu=0$), implying that $E^0=0=B^0$ for a cosmological background. With the previous definitions we decompose the electromagnetic field tensor as
\begin{equation}
 F^{\mu\nu}=E^\mu n^\nu-E^\nu n^\mu-\epsilon^{\mu\nu\rho\sigma}B_\rho n_\sigma\, .
\label{FaradayD}
\end{equation}
This allow us to  explicitly write the Maxwell equations \eqref{Maxwellcurved}  in terms of electric and magnetic fields  for a cosmological background by projecting them into time-like and space-like hypersurfaces. Substituting \eqref{FaradayD} into \eqref{Maxwellcurved}, and projecting it onto $n_\mu$ we find \cite{thorne,tajima,tarkenton}
\begin{equation}
\nabla\cdot{\bm E}=4\pi q n \Gamma\, .
\label{div1E}\end{equation}
Also, projecting Eq.~\eqref{Maxwellcurved} onto space-like hypersurfaces yields
\begin{equation}
\nabla\times{\bm B}=4\pi qn\Gamma\bm v+\frac{1}{a^3}\frac{\partial\left(a^3{\bm E}\right)}{\partial t}\, .
\label{div2E}\end{equation}
Similarly, one can define the dual electromagnetic tensor
\begin{eqnarray}
 F^{*\mu\nu}=\frac{1}{2}\epsilon^{\mu\nu\rho\tau}F_{\rho\tau}=B^\mu n^\nu-B^\nu n^\mu-\epsilon^{\mu\nu\rho\tau}E_\rho n_\tau\, ,
\end{eqnarray}
that satisfies ${F^{*\mu\nu}}_{;\nu}=0$ by its antisymmetry. When projected onto $n_\mu$,
we find the time-like component \cite{thorne,tajima,tarkenton}
\begin{equation}
 \nabla\cdot\bm B=0\, ,
\label{div1B}
\end{equation}
whereas the spacelike projection has the vectorial equivalent \cite{thorne,tajima,tarkenton}
\begin{equation}
\frac{1}{a^3}\frac{\partial\left(a^3\bm B\right)}{\partial t}=-\nabla\times\bm E\, .
\label{div2B}
\end{equation}
Eqs.~\eqref{div1E}, \eqref{div2E}, \eqref{div1B} and \eqref{div2B} correspond to the Maxwell equations for a cosmological plasma in an expanding universe.

For plasma dynamics, a similar decomposition can be performed on the antisymmetric unified tensor  \eqref{Mu} yielding the  generalized electric $\xi^\mu$ and magnetic $\Omega^\mu$ fields,
\begin{equation}
 \xi^\mu=n_\nu M^{\nu\mu}\, ,\qquad \Omega^\mu=\frac{1}{2}n_\rho \epsilon^{\rho\mu\sigma\tau}M_{\sigma\tau}\, ,
\end{equation}
that are both spacelike ($n_\mu \xi^\mu=0$ and $n_\mu \Omega^\mu=0$). Equivalently, $M^{\mu\nu}$ may be written as
\begin{equation}
 M^{\mu\nu}=\xi^\mu n^\nu-\xi^\nu n^\mu-\epsilon^{\mu\nu\rho\sigma}\Omega_\rho n_\sigma\, .
\label{Mdecomposicion}
\end{equation}


The $n_\mu$ projection will give the  (three-vector) generalized electric ${\bm \xi}$ and magnetic fields ${\bm \Omega}$ 
\begin{equation}
{\bm \xi}={\bm E}-\frac{m}{q a^2}\left[\nabla\left(f\Gamma\right)+\frac{\partial}{\partial t}\left(fa^2\Gamma {\bm v}\right)\right]\, ,
\label{GE}\end{equation}
\begin{equation}
{\bm \Omega}={\bm B}+\frac{m a^2}{q}\nabla\times\left(f\Gamma {\bm v}\right)\, ,
\label{GVO}\end{equation}
the curved spacetime generalization of the corresponding vector fields defined in Refs.~\cite{mah1,mah12}. We emphasize that the  generalized magnetic field ${\bm \Omega}$ allows an  interpretation as a generalized vorticity because  it is, indeed, the  curl of a potential  (${\bm \Omega}=\nabla\times {\cal A}$), where
\begin{equation}
{\cal A}={\bm A}+\frac{a^2 m f \Gamma}{q} {\bm v}\, ,
\label{GVOpotn}\end{equation}
and ${\bm A}$ is the vector potential of the electromagnetic field. From now, the names ``generalized magnetic fields" and ``generalized vorticity" are used interchangeably.

The generalized electric field \eqref{GE} and the generalized magnetic field \eqref{GVO}  are the key to writing the equation of motion \eqref{eqmo} in an insightful form. With previous definitions, Eq.~\eqref{eqmo} becomes
\begin{equation}
 \xi^\mu-a^2 \gamma_{ij} \xi^i v^j n^\mu+ n_\lambda \epsilon^{\lambda\mu\nu\rho}v_\nu \Omega_\rho=-\frac{T}{q\Gamma}\partial^\mu\sigma\, .
\label{eqmoGR}
\end{equation}
This is the covariant form of the equation of motion from where the  $3+1$ equations can be obtained by appropriated projections on the time-like and space-like hypersurfaces. The  $n^\mu$ projection  gives arise to the equation for energy conservation
\begin{equation}
a^2 {\bm v}\cdot {\bm \xi}=\frac{T}{q\Gamma}\frac{\partial\sigma}{\partial t}\, ,
\label{energyCon}\end{equation}
while the spacelike ${\gamma^\beta}_{\mu}$ projection  yields the vectorial momentum evolution equation
 \begin{equation}
{\bm \xi}+{\bm v}\times{\bm\Omega}=-\frac{T}{q\Gamma}\nabla\sigma\, .
\label{momentCon}
\end{equation}
Eqs.~\eqref{energyCon} and \eqref{momentCon} are equivalent to the usual $3+1$ plasma equations \cite{tajima,tarkenton} invoked in plasma literature. Notice that there exist effects of the interaction of the fluid with the  spacetime expansion hidden in the definition of the unified fields.

This unified magnetofluid approach leads us directly to the general vortical form of depicting the plasma dynamics. In this formalism the sources of general vorticity (where the magnetic field is just a component) are explicitly revealed. The vortical plasma dynamics can be completly described by using the antisymmetric properties of the  unified tensor $M^{\mu\nu}$. Its dual tensor follows the conservation equation
\begin{equation}\label{consdual}
\nabla_\nu M^{*\mu\nu}=0\, ,
\end{equation}
where $M^{*\mu\nu}=(1/2)\epsilon^{\mu\nu\alpha\beta} M_{\alpha\beta}$ (in analogy with the electromagnetic tensor). A $3+1$ decomposition of this equation provides  physical insights on vortical dynamics. The dual tensor,
\begin{equation}\label{Mdual}
M^{*\mu\nu}=\Omega^\mu n^\nu-\Omega^\nu n^\mu+\epsilon^{\mu\nu\alpha\beta} \xi_\alpha n_\beta\, ,
\end{equation}
on $3+1$ decomposition, leads to the equation for the timelike hypersurface 
\begin{equation}
\nabla\cdot {\bm \Omega}=0\, .
\end{equation}
This equation represent the generalization of the divergence-free nature of the magnetic field. On the other hand,  the spacelike projection of Eq.~\eqref{consdual},
\begin{equation}
\frac{\partial}{\partial t}\left(a^3 {\bm \Omega}\right)+a^3\nabla\times{\bm \xi}=0\, ,
\label{div2BM}
\end{equation}
represents the constraint linking the generalized electric and magnetic fields (Generlized Faraday law).

\section{Plasmas in classical perfect diamagnetism state- Magnetic field structures}

In formulating plasma dynamics beyond MHD, we noticed that if generalized vorticity were to replace the magnetic field, the MHD like vortical structure of the dynamics is fully retained. Since the velocity and magnetic fields are the measurables of interest, we will seek
self-consistent solutions for ${\bm v}$ and $\bm B$ for a specified thermodynamics (the current model does not evolve thermodynamics). A variety of such solutions for the special relativistic dynamics were worked out in Ref.~\cite{mahajanElectroVortical}. In this section, we will explore the appropriate translations of  some of these solutions in the context of cosmological plasmas in an expanding universe. 

An interesting (and exact) class of solutions, accessible to cosmological plasmas, belong to the general category of states that display Classical Perfect Diamagnetism (CPD) \cite{mahajanclassicaldiag}. In CPD states, the generalized vorticity is fully expelled  from the plasma interior. It is worthwhile to remark here that it is the  vanishing of the canonical vorticity that leads to the  London equation (implying Meissner Ochsenfeld effect) describing a standard superconductor.
 
Let us now derive the equations that  define the CPD state pertinent to the cosmological plasma. In its simplest manifestation in a  homentropic plasma ($\partial^\mu\sigma=0$), Eq.~\eqref{eqmo} allows the solution 
\begin{equation}
M^{\mu\nu}= 0\, ,\qquad F^{\mu\nu}=-\frac{m}{q}S^{\mu\nu}\, ,
\end{equation}
that has a vanishing generalized vorticty, ${\bm\Omega}=0$. More explicitly, this condition relates the magnetic and velocity fields [see Eq.~\eqref{GVO}],
\begin{equation}\label{SuperCondSta}
{\bm B}=-\frac{m a^2}{q}\nabla\times\left(f\Gamma{\bm v}\right)\, .
\end{equation}
Note that \eqref{SuperCondSta} and the Maxwell equations \eqref{div2E}- \eqref{div2B} form a self-consistent system for the magnetic and velocity fields. In fact, the combination yields  
\begin{equation}\label{SCS1}
\nabla^2{\bm B}=\frac{4\pi q^2 n}{m f a^2} {\bm B}+\frac{1}{a^3}\frac{\partial^2}{\partial t^2}\left(a^3{\bm B}\right)\, ,
\end{equation}
that can be solved for the magnetic field as long as the thermodynamic functions and the scale parameter  $a$ are specified. There are no  explicit external drives, relativistic or otherwise, needed to catapult the  system from a zero to finite magnetic field state.  

Unlike the standard equations associated with superconducting states, Eq.~\eqref{SCS1} is time dependent. An exactly solvable set emerges if we assume that the thermodynamic quantities ($n$ and $f$) are functions of time only (through the scale factor $a$). In such a case, 
 the ansatz, ${\bm B}({\bm x}, t)= {\bm b}({\bm x}) {\cal T}(t)/a(t)^3$ splits \eqref{SCS1} into two ordinary differential equations
\begin{equation}\label{SCS2}
\nabla^2{\bm b}=\lambda_p^2\, {\bm b}\, ,
\end{equation}
\begin{equation}\label{SCS30}
\frac{\partial^2 {\cal T}}{\partial t^2}+\omega_p^2\left(\frac{\widehat {n}}{a^2  \widehat{f}}-1\right) {\cal T}=0\, ,
\end{equation}
for the spatial and the temporal parts. The spatial part \eqref{SCS2} is, precisely, the London equation predicting the spatial decay of ${\bm b}$ on a collisionless  thermally corrected skin depth $\lambda_p={c}/{\omega_p}$, where  $\omega_p=\sqrt{4\pi q^2 n_0/(f_{0} m)}$ is the thermally corrected plasma frequency. For the temporal part of the magnetic field,  we have assumed  that $n=\widehat {n} n_0$, $f=\widehat {f}  f_0$, with $\widehat {n}$ and $\widehat {f}$ denoting the temporal variation of the profiles.

To explicitly evaluate the temporal behavior of the magnetic field, let us study the solution in the radiation dominated era when the scale factor increases as $a=a_0 t^{1/2}$ \cite{ryden}. Similarly, a hot plasma has a temperature that increases as $T\propto a^{-1}$, where its density is $n\propto T^3$. For high temperatures, $f\approx 4 k_B T/m$, and therefore  $\widehat {n} =a^{-3}$,  and $\widehat {f}=a^{-1}$. Thereby,  Eq.~\eqref{SCS30} becomes
\begin{equation}\label{SCS3}
\frac{\partial^2 {\cal T}}{\partial t^2}+\omega_p^2\left(\frac{1}{a_0^4  t^2}-1\right) {\cal T}=0\, ,
\end{equation}
which can be solved exactly in terms of Bessel functions. Since the era of interest for the current enquiry belongs to relatively smaller times ($a_0^2\,  t \ll 1$), the approximate but explicit solution
\begin{equation}
{\cal T}(t)\approx t^{\lambda/2}
\end{equation}
where
\begin{equation}
\lambda=1-\sqrt{1-\frac{4\omega_p^2}{a_0^4}}\, , \qquad 0<\lambda<1\, ,
\end{equation}
is more instructive. Notice that for real  $\lambda$ ($\omega_p<{a_0^2}/2$)  such a solution persists only for a fraction of the plasma time (inverse of the plasma frequency).

The magnetic field, described by Eqs.~\eqref{SCS2} and \eqref{SCS3} (and their solutions), though characterizing a CPD state, does not describe an equilibrium state; it evolves with the expansion of the universe.   However, the spatial configuration remains as usual, described by the London equation. Physically, the solution ${\bm B}({\bm x}, t)\approx {\bm b}({\bm x}) t^{(\lambda-3)/2}$, for the radiation-dominated era, describes a perfect diamagnetic state of the primordial plasma that expells the magnetic field from its inner region. As the Universe expand, and the time grows, this CPD state is more feasible to achieve, as the magnetic field decays faster in the plasma region.

\section{Magnetic field in classical super-diamagnetism state}

If the cosmological plasma were not homentropic (it must be isentropic in this model), it still allows the  equivalent of a superconducting state for a particular  class of velocity fields that satisfy \eqref{ClebschU}.  Such a state, called the  classical super-diamagnetic state (CSD), was analyzed for the special relativistic plasma in Ref.~\cite{mahajanElectroVortical}. We will now find the corresponding state for a cosmological plasmas in an expanding metric. For details the reader is referred to Ref.~\cite{mahajanElectroVortical}

For a non-homentropic plasma, the dynamics can be described by  Eq.~\eqref{eqmoUM}. Similar to previous case, the equation of motion Eq.~\eqref{eqmoUM} is identically solved if ${\cal M}^{\mu\nu}= 0$ which is insured if the generalized potential 
\begin{equation}\label{AmuCSD}
A^\mu+\frac{m}{q}\left(f-\frac{\sigma T}{m}\right)U^\mu=0\, .
\end{equation}
The curl of the vector part of \eqref{AmuCSD}  yields the CSD state, defined by (${\bar f}=f-\sigma T/m$)
\begin{equation}\label{SuperCondSta2}
{\bm B}=-\frac{m a^2}{q}\nabla\times\left({\bar f}\Gamma{\bm v}\right)\, .
\end{equation}
Combining \eqref{SuperCondSta2} with Maxwell equations leads (like the last section) to 
\begin{equation}\label{CSD1}
\nabla^2{\bm B}=\frac{4\pi q^2 n}{m {\bar f} a^2} {\bm B}+\frac{1}{a^3}\frac{\partial^2}{\partial t^2}\left(a^3{\bm B}\right)\, ,
\end{equation}
which is exactly the same as \eqref {SCS1} with $\bar f$ replacing $f$. Thereby, it can be solved and analyzed exactly in the same manner.  
Notice that the CPD [Eq.~\eqref {SuperCondSta}] and CSD [Eq.~\eqref{SuperCondSta2}] conditions differ in that the latter, in principle, allows variation in entropy (consistent with the isentropic condition). However for spatially constant thermodynamics, as invoked in previous sections, the isentropic condition forces a time independent entropy, and thus $\bar f$ temporally depends on the behavior of $f$ and temperature $T$.

\section{Discussion}

We  have worked out, in this paper, the cosmological version ( in an expanding universe) of the  self-consistent magnetic and velocity fields that can exist even in plasmas with vanishing generalized vorticies. Such ``superconducting" macroscopic states do not require any external seed mechanisms, such as Biermann battery \cite{massimo1, massimo2, naoz,kulsrud} or its generalizations. 

Both classes of previously known diamagnetic states (magnetic fields nonzero only over a skin depth), when translated into the cosmic context, yield fields that grow with the expansion as ${\cal T}\propto a^{\lambda}$ (while the scale factor goes up as $a\propto t^{1/2}$). The temporal growth of $B$ tends to compete with the dilution (as $a^{-3}$) caused by the cosmological expansion. These new states with initially growing fields (whose detailed nature is yet to be explored) add a brand new element towards advancing our understanding of the cosmological magnetic fields and flows.

We have demonstrated, using the  ElectroVortic formulation of the dynamics of relativistic plasmas in a time dependent metric, that  the growing magnetic and vorticity fields are automatic solutions in an expanding universe. This newly established characteristic can and does bring additional insights into the comprehension  of the ``origin" of primordial  magnetic fields in the Universe. This new set of solutions enriches the study of cosmological plasmas in general \cite{holcomb,tajimashibata}, and can provide highly relevant initial conditions for dynamo action \cite{pudritz,colgate,khanna2,rein,fleishman}.

\begin{acknowledgments}

The work of SMM was supported by USDOE Contract No.DE-- FG 03-96ER-54366.  

\end{acknowledgments}

\end{document}